\documentclass[DIV=calc,paper=a4,fontsize=11pt,twocolumn]{scrartcl} 

\usepackage[english]{babel}
\usepackage[protrusion=true,expansion=true]{microtype}
\usepackage{amsmath,amsfonts,amsthm}
\usepackage[final]{graphicx}
\usepackage{xcolor}
\usepackage[normal,small,hypcap,up,labelfont=bf,textfont=it]{caption}
\usepackage{epstopdf}
\usepackage{subfig}
\usepackage{booktabs}
\usepackage{fix-cm}
\usepackage{amssymb,amsfonts}
\usepackage{dsfont}
\usepackage{bbm}
\usepackage{pstricks}
\usepackage{cite}
\usepackage[utf8]{inputenc}
\usepackage[perpage,symbol*]{footmisc}
\usepackage[varg]{txfonts}
\usepackage{balance}
\usepackage{fancyhdr}
\PassOptionsToPackage{hyphens}{url}\usepackage[pdfencoding=auto,psdextra]{hyperref}
\usepackage{bookmark}
\usepackage{verbatim}
\usepackage{fontenc}
\usepackage{cuted}
\usepackage{braket}

\usepackage{bm}
\usepackage{mathrsfs}

\theoremstyle{definition}

\theoremstyle{plain}

\DeclareCaptionFont{mycolor}{\color[HTML]{000000}}
\usepackage{sectsty}													
\allsectionsfont{
\color[HTML]{31ADF3}\usefont{OT1}{phv}{b}{n}
}

\sectionfont{
\color[HTML]{31ADF3}\usefont{OT1}{phv}{b}{n}
}

\usepackage{fancyhdr}												
\usepackage{lettrine}
\newcommand{\initial}[1]{%
\lettrine[lines=3,lhang=0.3,nindent=0em]{
\color[HTML]{31ADF3}
{\textsf{#1}}}{}}

\usepackage{titling}															

\newcommand{\HorRule}{\color[HTML]{31ADF3}
\rule{\linewidth}{1pt}%
}

\pretitle{\vspace{-30pt} \begin{flushleft} \HorRule
\fontsize{34}{34} \usefont{OT1}{phv}{b}{n} \color[HTML]{31ADF3} \selectfont
}
\title{Hanbury Brown--Twiss Effect with Wave Packets}					
\posttitle{\par\end{flushleft}\vskip 0.5em}

\preauthor{\begin{flushleft}\large \lineskip 0.5em \usefont{OT1}{phv}{b}{sl} \color[HTML]{31ADF3}}
\author{Tabish Qureshi$^{~\mathsf{1}}$ \& Ushba Rizwan$^{~\mathsf{2}}$\\[8pt]}											
\postauthor{\footnotesize \usefont{OT1}{phv}{m}{sl} \color[HTML]{000000}
$^{\mathsf{1}}$ Centre for Theoretical Physics, Jamia Millia Islamia, New Delhi,
India. E-mail: \href{mailto:tabish@ctp-jamia.res.in}{tabish@ctp-jamia.res.in}\\
$^{\mathsf{2}}$ Department of Physics, Jamia Millia Islamia, New Delhi, India. E-mail: \href{mailto:93.ushba@gmail.com}{93.ushba@gmail.com}\\[10pt]		
\scriptsize\usefont{OT1}{phv}{m}{n} \color[HTML]{31ADF3}{\textbf{Editors: \emph{Radu Ionicioiu} \& \emph{Danko Georgiev}} }\\[5pt]
\color[HTML]{000000}{Article history: Submitted on October 27, 2017;  Accepted on November 28, 2017; Published on November 30, 2017.}
\par\end{flushleft}\HorRule}

\date{}																				

\begin{document}
\maketitle
\thispagestyle{fancy} 			
\initial{T}\textbf{he Hanbury Brown--Twiss (HBT) effect, at the quantum level, is essentially an interference of one particle with another, as opposed to interference of a particle with itself. Conventional treatments of identical particles encounter difficulties while dealing with entanglement. A recently introduced label-free approach to indistinguishable particles is described, and is used to analyze the HBT effect. Quantum wave-packets have been used to provide a better understanding of the quantum interpretation of the HBT effect. The effect is demonstrated for two independent particles governed by Bose--Einstein or Fermi--Dirac statistics. The HBT effect is also analyzed for pairs of entangled particles. Surprisingly, entanglement has almost no effect on the interference seen in the HBT effect. In the light of the results, an old quantum optics experiment is reanalyzed, and it is argued that the interference seen in that experiment is not a consequence of non-local correlations between the photons, as is commonly believed.\\ Quanta 2017; 6: 61--69.}

\begin{figure}[b!]
\rule{245 pt}{0.5 pt}\\[3pt]
\raisebox{-0.2\height}{\includegraphics[width=5mm]{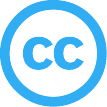}}\raisebox{-0.2\height}{\includegraphics[width=5mm]{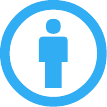}}
\footnotesize{This is an open access article distributed under the terms of the Creative Commons Attribution License \href{http://creativecommons.org/licenses/by/3.0/}{CC-BY-3.0}, which permits unrestricted use, distribution, and reproduction in any medium, provided the original author and source are credited.}
\end{figure}

\section{Introduction}

\begin{figure*}[t!]
\centering
\includegraphics[width=168mm]{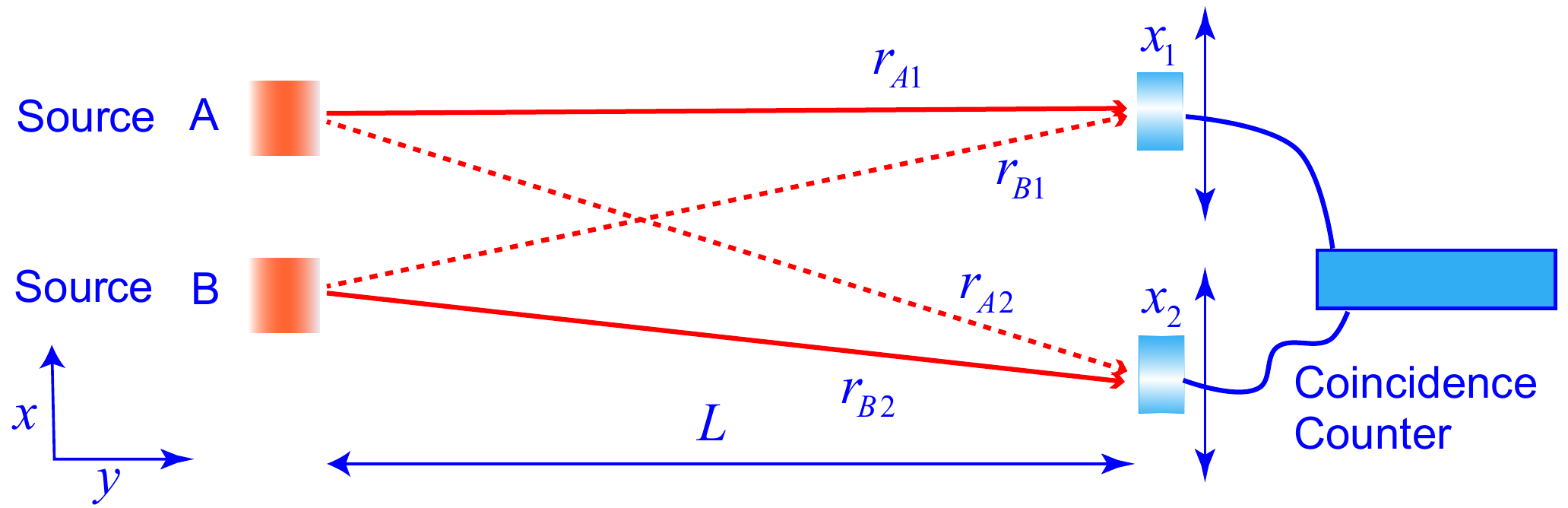}
\caption{A schematic diagram for a two source interference, and also
for a intensity correlation at two points. Sources A and B are stationary
while the detectors at $x_1$ and $x_2$ are movable.}
\label{schematic}
\end{figure*}

Interference of waves is a very old and well studied subject. Waves 
emanating from two sources give rise to interference, provided that
there is a coherence in the phases of the two. This can be understood with
the following argument. Suppose there are two sources~A~and~B, waves
emerging from which fall on a distant screen. At a point~$x_1$ on the
screen, the contribution of the two classical waves can be written as
\begin{equation}
{\cal E}(x_1) = \alpha e^{\imath kr_{A1}} + \beta e^{\imath kr_{B1}-\imath \phi},
\end{equation}
where $k$ is the wave-vector of the two waves, $r_{A1}, r_{B1}$
are the displacements as shown in Figure~\ref{schematic}, and $\phi$ is the
difference in the phases of the two sources. Assuming $\alpha,\beta$
to be real, the intensity at the point $x_1$ is then given by
\begin{equation}
I(x_1) \equiv |{\cal E}(x_1)|^2 = \alpha^2 + \beta^2 + 
2\alpha\beta \cos\left[{k(r_{A1}-r_{B1})+\phi}\right].
\end{equation}
The  above expression represents interference between the wave from A and B, at point
$x_1$ on the screen. One can see that if the phase difference between the
two waves, $\phi$, fluctuates randomly with time, which amounts to
integrating the above over $\phi$, the interference will be washed away.
\begin{equation}
\langle I(x_1)\rangle \approx \alpha^2 + \beta^2 . 
\end{equation}
Thus two independent, incoherent sources of waves do not give rise to
interference.
Hanbury Brown and Twiss carried out an experiment with radio waves which
involved correlating intensities at {\em two points} on the screen, coming
from two independent sources \cite{hbt0}. Their experiment can be easily
understood from the preceding example. An intensity correlation
at two points $x_1$ and $x_2$ on the screen can be written as
\begin{eqnarray}
I(x_1)I(x_2) &=& |{\cal E}(x_1)|^2 |{\cal E}(x_2)|^2 \nonumber\\
&=& \left[\alpha^2 + \beta^2 + 2\alpha\beta \cos({k(r_{A1}-r_{B1})+\phi})\right]\nonumber\\
&&\left[\alpha^2 + \beta^2 + 2\alpha\beta \cos({k(r_{A2}-r_{B2})+\phi})\right] \nonumber\\
\end{eqnarray}
After averaging over $\phi$, the $\phi$-dependent terms drop out, and
one is left with
\begin{eqnarray}
\langle I(x_1)I(x_2)\rangle &\approx& (\alpha^2 + \beta^2)^2\nonumber\\
&&+ 2(\alpha\beta)^2 \cos(k(r_{A1}-r_{B1}-r_{A2}+r_{B2})). \nonumber\\
\label{chbt}
\end{eqnarray}
Surprisingly, although individual intensities do not show any oscillation,
there is interference between the intensities at two different points on
the screen. This is what is called the Hanbury Brown--Twiss (HBT) effect.
Random fluctuation of phases of the independent source has no effect on
the oscillations seen in the intensity correlations.
Notice that the visibility \cite{born} of the above interference pattern is 
$\frac{2(\alpha\beta)^2}{(\alpha^2 + \beta^2)^2}$ which is bounded from
above by $1/2$.

The HBT effect was explained using classical waves, but when the same
experiment was  proposed using light, there was a lot of skepticism, and
people thought it would not work. The reason was that the HBT effect
arose from two different waves interfering with each other, and many
people believed that a photon interferes only with itself, two photons
never interfere with each other \cite[p.~9]{dirac}.
Hanbury Brown and Twiss carried out the experiment with light and
demonstrated the HBT effect \cite{hbt}.  The first quantum understanding
of the HBT effect for light was given by Fano \cite{fano}. 
The HBT effect has now been demonstrated for quantum light \cite{hbtlight1,hbtlight2}.
Not only that, it has also been demonstrated with massive particles,
both of bosonic \cite{hbtatom1,hbtatom2,hbtatom3} and fermionic \cite{hbtelectrons} nature.

The HBT effect for massive particles is intriguing, as it involves
interference between two different particles. Particles are seen to be
bunched together without any interaction between them. It is a purely
quantum mechanical effect, and the classical
wave explanation is only a part of the story. Full treatment of the HBT
effect can of course be done using quantum many-body theory, as the particles
are treated as being indistinguishable there. However, that maybe an
overkill, as it is essentially a two particle effect. In the following
we treat two massive particles as traveling wave-packets, and use them
to analyze the HBT effect.

\section{Label-free approach to identical particles}
\label{identical}

For a quantum treatment of the HBT effect, it is essential to take the
indistinguishability of the particles into account. However, treatment
of identical particles in quantum mechanics, particularly those involving
entanglement, has been an issue which has been
much debated \cite{deb1,deb2,deb3,deb5,deb6,deb7,deb8,deb9,deb10, 
deb11,deb12,deb13,deb14,deb15,deb16,deb17,deb18,deb19}.
The problem can be seen in the very basic symmetric or anti-symmetric
wavefunctions that are usually written, e.g.
$\phi_A(x_1)\phi_B(x_2) \pm \phi_B(x_1)\phi_A(x_2)$. Although the particles
are assumed to be indistinguishable, we put labels 1 and 2 on them.
In addition, since the above state is not separable, the question arises,
whether it is an entangled state?
The problem of labeling, of course, disappears if one uses the second
quantization approach \cite{9qm}.
However, the problems in dealing with entangled states do
not go away just by using second quantization.
Recently a new state-based approach has been introduced which does not
label the particles, and appears to address the entanglement related issues
satisfactorily \cite{franco}. In the following, we will explain this
approach and use it to analyze HBT experiment with quantum particles.

In this new approach, the basic assumption is that in dealing with more than
one identical particles, the particles cannot be individually addressed.
This is in accordance with the quantum theory. The combined state of two
particles, which may consist of single-particle states, is treated as
a holistic indivisible entity \cite{franco}. One cannot ask for the form
of this state. What one can ask, is what the two particle probabiliity
amplitude of finding the two particles in two different states is. For
example, if one particle is in the single-particle state $|\psi\rangle$
and one in the state $|\phi\rangle$, the two-particle state is represented
as $|\phi,\psi\rangle$. One may not talk about the ``form'' of this state
in terms of single-particle states $|\psi\rangle$ and $|\phi\rangle$.
However, the probability amplitude of finding one particle in state
$|\alpha\rangle$ and the other in state $|\beta\rangle$, i.e., in the
combined state $|\alpha,\beta\rangle$, is given by
\begin{equation}
\langle\alpha,\beta|\phi,\psi\rangle = \langle\alpha|\phi\rangle
\langle\beta|\psi\rangle + \eta \langle\alpha|\psi\rangle
\langle\beta|\phi\rangle,
\label{statedef}
\end{equation}
where $\eta^2 = 1$. One can check that the state $|\phi,\psi\rangle$ is not
normalized simply by replacing $|\alpha,\beta\rangle$ by $|\phi,\psi\rangle$
in the above equation. The state $|\phi,\psi\rangle$ has to be multiplied
with $\frac{1}{\sqrt{1+\eta|\langle\phi|\psi\rangle|^2}}$ in order to
normalize it. The probability amplitude of finding the particles at positions
$x_1$ and $x_2$, i.e., in the state $|x_1,x_2\rangle$ is given by
\begin{equation}
\langle x_1, x_2|\phi,\psi\rangle = \langle x_1|\phi\rangle
\langle x_2|\psi\rangle + \eta \langle x_1|\psi\rangle
\langle x_2|\phi\rangle,
\end{equation}
which is the familiar symmetric or antisymmetric two-particle wavefunction.
The difference is that, in this case 1 and 2 are not labels of particles,
but corresponds to the two positions of a joint measurement.

A one-particle operator $\hat{A}$ acts on the two-particle state in the
following way:
\begin{equation}
\hat{A}|\phi,\psi\rangle = |\hat{A}\phi,\psi\rangle + |\phi,\hat{A}\psi\rangle.
\label{opdef}
\end{equation}
The expectation value of a one-particle operator, using (\ref{statedef}) and
(\ref{opdef}), is given by
\begin{eqnarray}
\langle\hat{A}\rangle &=& \langle\phi,\psi|\hat{A}|\phi,\psi\rangle\nonumber\\
&=& \langle\phi|\hat{A}\phi\rangle + \langle\psi|\hat{A}\psi\rangle\nonumber\\
&&+\eta\left(\langle\phi|\psi\rangle\langle\psi|\hat{A}\phi\rangle
+ \langle\psi|\phi\rangle\langle\phi|\hat{A}\psi\rangle\right),
\end{eqnarray}
which agrees with the expression of the conventional analysis.

\section{Independent particles}
\label{indep}

Let there be two particles described by two wave packets, traveling along
y-axis. The two particles emerge from two sources localized at positions
$x_0$ and $-x_0$. We described the two particles by two Gaussian
wave packets of width $\epsilon$ each, localized at $x_0$ and
$-x_0$, denoted by $|\phi_A\rangle$ and $|\phi_B\rangle$, respectively.
Since the particles are indistinguishable, the combined initial state of
the two particles can be written as
\begin{equation}
|\psi(0)\rangle = |\phi_A,\phi_B\rangle
\end{equation}
This satisfies the essential requirement for HBT effect, that the particles be identical,
in the quantum sense.
The conventional two-particle wavefunction is then just the probability
amplitude of finding one particle at $x_1$ and one at $x_2$, and can be
written down using (\ref{statedef}) as follows: 
\begin{eqnarray}
\psi(x_1,x_2,0) &=& \langle x_1, x_2|\phi_A,\phi_B\rangle\nonumber\\
&=& \langle x_1|\phi_A\rangle
\langle x_2|\phi_B\rangle + \eta \langle x_1|\phi_B\rangle
\langle x_2|\phi_A\rangle\nonumber\\
 &=&{1\over\sqrt{\pi}\epsilon}\left(e^{-(x_1-x_0)^2\over\epsilon^2}e^{-(x_2+x_0)^2\over\epsilon^2}\right.\nonumber\\
&&\left. + \eta e^{-(x_1+x_0)^2\over\epsilon^2}e^{-(x_2-x_0)^2\over\epsilon^2}\right)
\label{psi0}
\end{eqnarray}
where $\eta = \pm 1$.
The last line in the above equation
specifies the Gaussian form of the states emerging from the sources A and B,
namely, $\phi_A(x)= \exp{-(x-x_0)^2\over\epsilon^2}$ and
$\phi_B(x)= \exp{-(x+x_0)^2\over\epsilon^2}$.
For bosonic particles, the wavefunction should be
symmetric, and $\eta$ should be 1. For fermions, the two-particle wavefunction
should be antisymmetric, requiring $\eta$ to be $-1$.
We assume that the particles are traveling along the positive $y$-direction
with a constant velocity $v_0$. For simplicity we ignore the explicit 
time evolution along the y-axis, and assume that evolution for a time
$t'$ just transports the wave packets by a distance $l = v_0t'$. The 
dispersion of the wave packets along the transverse $x$-direction is more
interesting and may give rise to interference between wave packets.

Notice that if one of the two sources produces a wave packet with an
additional phase factor, say $e^{\imath \phi}$, that phase factor will present
in the both the terms in (\ref{psi0}), and can be pulled out, becoming
irrelevant.

The Hamiltonian governing the time evolution is that of two free particles.
The two-particle eigenstate will simply be $|p,p'\rangle$ which means one
particle has momentum $p$ and the other $p'$. One can thus write
\begin{equation}
\hat{H}|p,p'\rangle = \left({p^2\over 2m} + {p'^2\over 2m}\right)|p,p'\rangle
\label{hfree}
\end{equation}
After traveling for a time $t$, the particles reach the screen.
Time evolution of the initial state $|\phi_A, \phi_B\rangle$ can be worked
out by introducing a complete set of two-particle momentum eigenstates:
\begin{eqnarray}
|\psi(t)\rangle &=& e^{-\imath \hat{H}t/\hbar} |\phi_A, \phi_B\rangle\nonumber\\
&=& e^{-\imath \hat{H}t/\hbar}\sum_{p,p'}|p,p'\rangle\langle p,p'||\phi_A, \phi_B\rangle \nonumber\\
&=& \sum_{p,p'} e^{-\imath \frac{(p^2+p'^2)t}{2m\hbar}}|p,p'\rangle \nonumber\\
&&\left[\langle p|\phi_A\rangle\langle p'|\phi_B\rangle\right.
\left.+ \eta \langle p|\phi_B\rangle\langle p'|\phi_A\rangle\right] \nonumber\\
&=& \sum_{p,p'} |p,p'\rangle \nonumber\\
&&\left[\langle p|\phi_A(t)\rangle\langle p'|\phi_B(t)\rangle\right.
\left.+ \eta \langle p|\phi_B(t)\rangle\langle p'|\phi_A(t)\rangle\right] \nonumber\\
&=& |\phi_A(t), \phi_B(t)\rangle,
\end{eqnarray}
where
\begin{eqnarray}
\langle p|\phi_A(t)\rangle =  e^{-\frac{\imath p^2t}{2m\hbar}}\langle p|\phi_A\rangle ,~~~
\langle p'|\phi_B(t)\rangle =  e^{-\frac{\imath p'^2t}{2m\hbar}}\langle p|\phi_B\rangle. \nonumber\\
\end{eqnarray}
From the momentum representation of $|\phi_A\rangle$ and $|\phi_B\rangle$,
the position representation can be evaluated.  The amplitude of finding the
particles at positions $x_1$ and $x_2$ then works out to be
\begin{eqnarray}
\psi(x_1,x_2,t) &=& \alpha
\left(e^{-(x_1-x_0)^2\over\epsilon^2+\imath \Delta}
e^{-(x_2+x_0)^2\over\epsilon^2+\imath \Delta}\right.
\left. + \eta e^{-(x_1+x_0)^2\over\epsilon^2+\imath \Delta}
e^{-(x_2-x_0)^2\over\epsilon^2+\imath \Delta}\right),\nonumber\\
\end{eqnarray}
where $\Delta\equiv 2\hbar t/m$, and $\alpha=\tfrac{1}{\sqrt{\pi(\epsilon+{\imath \Delta/\epsilon})}}$.
For simplicity we introduce the notation $\psi_t = \psi(x_1,x_2,t)$.
The joint probability density of finding the particles at $x_1$ and $x_2$ is
given by
\begin{eqnarray}
|\psi_t|^2 &=&{1\over \pi\sigma^2}
\left(e^{-2\epsilon^2(x_1-x_0)^2\over\epsilon^4+\Delta^2}
e^{-2\epsilon^2(x_2+x_0)^2\over\epsilon^4+\Delta^2}\right.\nonumber\\
&&\left. + e^{-2\epsilon^2(x_1+x_0)^2\over\epsilon^4+\Delta^2}
e^{-2\epsilon^2(x_2-x_0)^2\over\epsilon^4+\Delta^2}\right.\nonumber\\
&&\left. + \eta e^{-(x_1+x_0)^2\over\epsilon^2+\imath \Delta}
e^{-(x_2-x_0)^2\over\epsilon^2+\imath \Delta}
e^{-(x_1-x_0)^2\over\epsilon^2-\imath \Delta}
e^{-(x_2+x_0)^2\over\epsilon^2-\imath \Delta}\right.\nonumber\\
&&\left. + \eta e^{-(x_1-x_0)^2\over\epsilon^2+\imath \Delta}
e^{-(x_2+x_0)^2\over\epsilon^2+\imath \Delta}
e^{-(x_1+x_0)^2\over\epsilon^2-\imath \Delta}
e^{-(x_2-x_0)^2\over\epsilon^2-\imath \Delta}\right),
\label{qhbt0}
\end{eqnarray}
where $\sigma^2 = \epsilon^2+\Delta^2/\epsilon^2$.
The above simplifies to
\begin{eqnarray}
|\psi_t|^2 &=&{2\over \pi\sigma^2}
e^{-2\epsilon^2(x_1^2+x_2^2+2x_0^2)\over\epsilon^4+\Delta^2}
\cosh\left[{4\epsilon^2(x_1-x_2)x_0\over\epsilon^4+\Delta^2}\right]\nonumber\\
&&\left(1 + \eta\frac{\cos\left({4\Delta(x_1-x_2)x_0\over\epsilon^4+\Delta^2}\right)}
{\cosh\left({4\epsilon^2(x_1-x_2)x_0\over\epsilon^4+\Delta^2}\right)}\right).
\label{qhbt}
\end{eqnarray}
The right-hand side of Eq.~(\ref{qhbt}) represents an interference pattern in the joint probability
of detection of the two particles, with respect to the distance between the
two positions $x_1$ and $x_2$ on the screen. For $\eta=1$, it constitutes the
HBT effect for massive particles, which obey Bose statistics. The same result
will also apply to independent photons with the proviso that
$\Delta = \lambda L/\pi$, where $\lambda$ is the  wave-length of the photons
and $L$ is the distance traveled by them in time $t$. In fact, the relation
$\Delta = \lambda L/\pi$ may also be used for massive particles in which
case $\lambda$ will represent the de Broglie wavelength of a particle.
It is easy to check that, for $|x_1-x_2|, x_0 \ll r_{A1}+r_{A2}, r_{B1}+r_{B2}$,
the cosine term in (\ref{chbt}) reduces to $\cos(4x_0(x_1-x_2)/\Delta)$, which
is the same as the cosine term in (\ref{qhbt}), provided that
$\epsilon^4 \ll \Delta^2$.

At this stage it may be useful to understand the physical meaning of various
terms in the joint probability distribution (\ref{qhbt0}).
In our usual
classical way of thinking, we imagine that there is a possibility of the
particle from source A reaching~$x_1$ and that from source B reaching~$x_2$.
The solids lines in Figure~\ref{schematic} represents this possibility and
the first term in (\ref{qhbt0}) represents its probability.
There is also the possibility of the particle from source A reaching~$x_2$
and that from source B reaching~$x_1$. The dashed lines in Figure~\ref{schematic} 
represent this possibility and the second term in (\ref{qhbt0})
represents its probability.
However, in quantum mechanics particles are not localized objects. In fact
they are capable of demonstrating wave and particle natures in different
situations. They should rather be called {\em quantons} \cite{quanton}.
Quantons may be visualized as fuzzy objects, sometime being spreadout like
a wave, and sometimes localized as particles. They can very well interfere
with themselves. If two independent quantons
can interfere with each other, there is also the possibility of 
quantons from A and B partially contributing to the quanton detected
at $x_1$, and at $x_2$. Since the quantons are identical, we have no way
of knowing which source they came from. In Figure \ref{schematic}, the
solid line $r_{A1}$ and the dotted line $r_{B1}$ represent the possibility of
A and B contributing to the quantons
reaching $x_1$, and the solid line $r_{B2}$ and the dotted line $r_{A2}$ represent
the possibility of B and A
contributing to the quantons reaching $x_2$.  The last two terms in 
(\ref{qhbt0}) represent this possibility. One can see that if the quantons
are not identical, the last two terms would not be there. This argument is
in agreement with the fact that the HBT effect cannot be seen for particles
which are not identical.

The last two terms can also be interpreted as describing interference
between the processes represented by the solid and the dashed lines. For certain
values of $x_1, x_2$ it might so happen that the solid lines process and
the dashed lines process destructively interfere. In that case, there will
be no simultaneous detection of quantons at all.

The visibility of this interference pattern is 
${1\over\cosh\left({4\epsilon^2(x_1-x_2)x_0\over\epsilon^4+\Delta^2}\right)}$,
which is bounded from above by 1. Contrast this with
the classical HBT effect described by (\ref{chbt}), where the visibility
cannot be greater than 1/2. This implies that there are certain distances
between the detectors 1 and 2 for which the probability of detecting
particles simultaneously is zero! The probability of detecting two particles
very close to each other is enhanced. This can be interpreted as a 
{\em bunching effect}. Remarkably, the particles tend to get bunched together
even when there is no interaction between them. This is a purely quantum
mechanical effect and has been experimentally observed in photons
\cite{hbtlight1,hbtlight2} as well as massive particles \cite{hbtatom1,hbtatom2,hbtatom3}.

For $\eta=-1$, the relation (\ref{qhbt}) implies that the probability
of simultaneously detecting two particles very close to each other is
nearly zero! Even when the particles strike the screen at random, there
is a certain probability of two of them hitting the screen very close to
each other. In the case $\eta=-1$, the probability of landing very close
to each other is even smaller than this random chance. It appears as if
the particles are repelling each other. This is what is called
{\em anti-bunching effect} and has been observed for particles following
Fermi--Dirac statistics, e.g., electrons \cite{hbtelectrons}.

\section{Entangled particles}
\label{entangled}

Let us now investigate the scenario where the particles are entangled.
There are certain sources of photons which generate photons in entangled
pairs. Entanglement manifests itself in strong quantum correlations between
the two particles. To our knowledge, the effect of entanglement on the HBT
effect has not been quantified.
Einstein, Podolsky and Rosen (EPR) first drew attention to a
momentum entangled state of two particles\cite{epr}
\begin{equation}
\Psi_{\textrm{EPR}}(x_1,x_2) = \!\int_{-\infty}^\infty 
e^{{\imath px_1\over\hbar}} e^{-\imath  px_2\over\hbar}
e^{-\imath 2x_0p\over\hbar} ~dp, \label{epr}
\end{equation}
which can be written in Dirac notation as
\begin{equation}
|\Psi_{\textrm{EPR}}\rangle = \!\int_{-\infty}^\infty
|p\rangle_1 |-p\rangle_2
e^{-\imath 2x_0p\over\hbar} dp, \label{eprdirac}
\end{equation}
where labels 1 and 2 refer to particle 1 and 2, respectively.
This state is for distinguishable particles. If one were to write an EPR
state for identical particles, in our label-free approach, it would be the
following
\begin{equation}
|\Psi_{\textrm{EPR}}^{\textrm{ident}}\rangle = \!\int_{-\infty}^\infty
|p,-p\rangle e^{-\imath 2x_0p\over\hbar} dp. \label{eprident}
\end{equation}
The amplitude of finding the particles at $x_1$ and $x_2$ is
given by
\begin{eqnarray}
\langle x_1,x_2|\Psi_{\textrm{EPR}}^{\textrm{ident}}\rangle
&=& \!\int_{-\infty}^\infty
\left[\langle x_1|p\rangle\langle x_2|-p\rangle\right.\nonumber\\
&&\left.+\eta \langle x_1|-p\rangle\langle x_2|p\rangle\right]
e^{-\imath 2x_0p\over\hbar} dp, \nonumber\\
\end{eqnarray}
which will be just the symmetrized or antisymmetrized form of (\ref{epr}).

The problem with the EPR state (\ref{epr}) is that it cannot be normalized,
and also it does not describe particles with varying degree of
entanglement. To address these shortcomings, we introduce a
{\em generalized EPR state} for identical particles
\begin{equation}
|\Psi\rangle = C\!\int_{-\infty}^\infty  \int_{-\infty}^\infty 
|q+p,q-p\rangle e^{-\imath 2x_0p\over\hbar}e^{-{p^2\over\hbar^2\sigma^2}}
e^{-{q^2\Omega^2\over\hbar^2}} ~dp~dq,
\label{geprd}
\end{equation}
where $q+p$ and $q-p$ label single-particle momentum eigenstates,
$C$ is a normalization constant, and $\sigma,\Omega$ are certain
parameters.
In the limit $\sigma,\Omega\to\infty$ the state (\ref{geprd})
reduces to the EPR state (\ref{eprident}), if $-2x_0$ here is identified with
$x_0$ in the EPR state.

The two-particle amplitude of finding them at $x_1$ and $x_2$ is given by
\begin{eqnarray}
\Psi(x_1,x_2) &=& \sqrt{ {\sigma\over \pi\Omega}}\left(
 e^{-(x_1-x_2-2x_0)^2\sigma^2} e^{-(x_1+x_2)^2/4\Omega^2}\right.\nonumber\\
&&\left. + \eta e^{-(x_1-x_2+2x_0)^2\sigma^2} e^{-(x_1+x_2)^2/4\Omega^2}\right).
\label{sgepr}
\end{eqnarray}
The state (\ref{sgepr}) is an extended version of the generalized EPR
state introduced earlier \cite{tqajp}.
It is straightforward to show that $\Omega$ and $\sigma$ quantify the position
and momentum spread of the particles in the $x$-direction. 
The interesting thing about this state is that for 
$2\Omega=1/\sigma=\epsilon\sqrt{2}$, it is no longer entangled and reduces
exactly to the symmetric state (\ref{psi0}) studied in the last section,
which is a symmetrized or anti-symmetrized product of two Gaussians
centered at $x_0$ and $-x_0$. 
So the entangled state is essentially two shifted Gaussians entangled
with each other.  
The state (\ref{sgepr}) is symmetric under the interchange of the two particles,
thus describing bosonic particles.

The stage is now set to study HBT effect with two entangled particles,
described by the state (\ref{sgepr}). The two particles travel in the
$y$-direction for a time $t$ before reaching the screen. During this time,
the states evolves in transverse $x$-direction too. As done in the last
section, we ignore the time evolution in the $y$-direction, and only
consider the evolution in the $x$-direction. If one is dealing with
photons, one can use an alternative wave-packet evolution \cite{tqsheeba}.
The state of the two particles, on reaching the screen (or detectors),
is given by
\begin{eqnarray}
\Psi(x_1,x_2,t) &=& C_t
 e^{\left[{-{(x_1+x_2)^2\over 4\Omega^2+\imath \delta}}\right]}
\left( \exp\left[-{(x_1-x_2-2x_0)^2\over 1/\sigma^2+\imath \delta}\right]\right.\nonumber\\
&&\left.+ \eta\exp\left[-{(x_1-x_2+2x_0)^2\over 1/\sigma^2+\imath \delta}\right] \right),
\end{eqnarray}
where $C_t = \sqrt{1\over\pi}\left[\left\{\Omega^2+\left({\lambda L\over 2\pi\Omega}\right)^2\right\}
\left\{{1\over\sigma^2}+\left({2\sigma\lambda L\over\pi}\right)^2\right\}\right]^{-1/4}$,
\mbox{$\delta = 4\hbar t/m = 2\lambda L/\pi$}
and $L$ is the distance in the \mbox{$y$-direction}, traveled by the particles during
time $t$.
\begin{figure*}[t!]
\centering
\includegraphics[width=168mm]{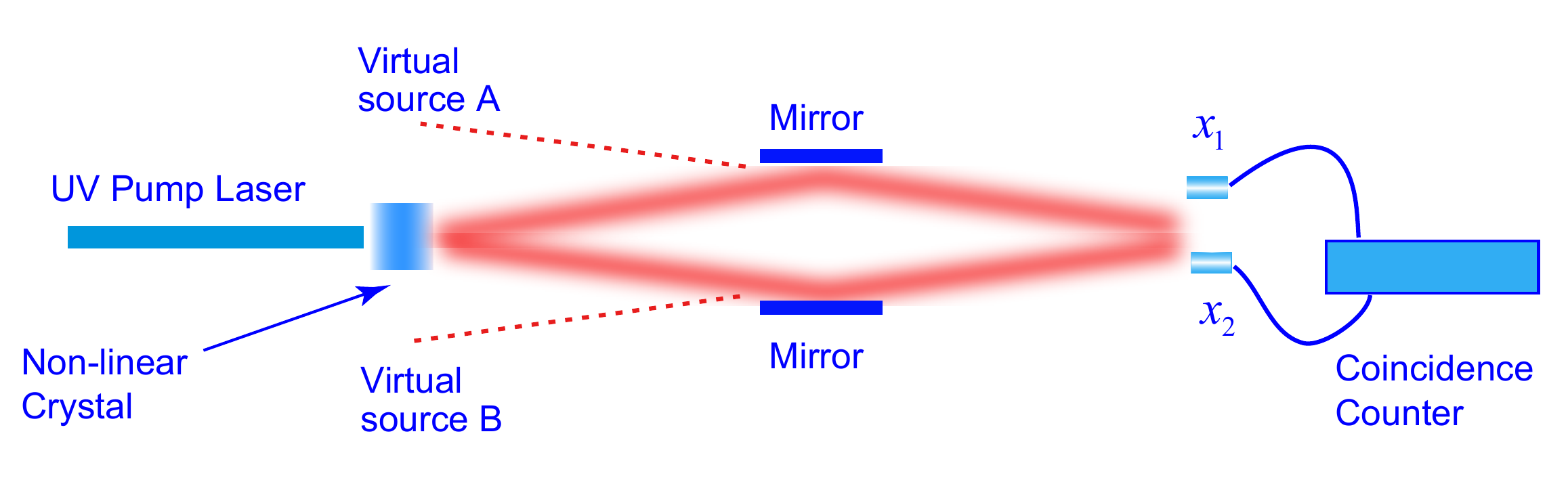}
\caption{A schematic diagram of the Ghosh--Mandel experiment \cite{ghoshmandel}.
A UV pump laser falls on a non-linear crystal and produces pairs of entangled
photons via spontaneous parametric down-conversion (SPDC), traveling in different
directions. These photons are deflected by two mirrors and recombined where
two movable detectors detect them in coincidence.}
\label{ghoshmandel}
\end{figure*}

The probability density of joint detection of particles at $x_1$ and $x_2$
can now be calculated, and is given by
\begin{eqnarray}
|\Psi(x_1,x_2,t)|^2 &=& |C_t|^2
e^{-8\Omega^2(x_1+x_2)^2\over 16\Omega^4+\delta^2}
e^{-2((x_1-x_2)^2+2x_0^2)/\sigma^2\over 1/\sigma^4+\delta^2}\nonumber\\
&&\cosh\left[{8(x_1-x_2)x_0/\sigma^2\over 1/\sigma^4+\delta^2}\right]\nonumber\\
&&\left(1 + \eta\frac{\cos\left({8\delta(x_1-x_2)x_0\over 1/\sigma^4+\delta^2}\right)}
{\cosh\left[{8(x_1-x_2)x_0/\sigma^2\over 1/\sigma^4+\delta^2}\right]}
\right).
\label{ehbt}
\end{eqnarray}
The above expression closely resembles (\ref{qhbt}) derived for particles
which are not entangled. Let us explore it in the situation when the
entanglement between the two particles is strong. From the Gaussian in 
$(x_1-x_2)$ in (\ref{sgepr}), one can see that as $\sigma$ increases, the
Gaussian narrows, and $(x_1-x_2-2x_0)$ becomes more localized, which 
implies stronger position correlation between detected particles.
Stronger position correlation implies stronger entanglement.
Thus, entanglement is strong when $\sigma$ is large
and one can safely assume $1/\sigma^4 \ll \delta^2$. In this limit, the
cosine term becomes $\cos[8(x_1-x_2)x_0/\delta] = \cos[4(x_1-x_2)x_0/\Delta]$,
which means that the interference fringe width is the same as in the
case of independent particles and also in the classical case. The Gaussian
terms in (\ref{ehbt}) assume the approximate form
$$e^{-2(x_1+x_2)^2\over 4\Omega^2+\delta^2/4\Omega^2}
e^{-2((x_1-x_2)^2+2x_0^2)\over \sigma^2\delta^2}.$$
These Gaussians represent
broad profiles both in $x_1+x_2$ and $x_1-x_2$. Therefore it appears that
entanglement does not cause any suppression of HBT effect, and remains
almost as it is for independent particles.

In the opposite limit, i.e., when the entanglement goes to zero, one 
can write $4\Omega^2 = 1/\sigma^2 \equiv 2\epsilon^2$ (say). In this limit,
(\ref{ehbt}) exactly reduces to (\ref{qhbt}), as expected.

\section{The Ghosh--Mandel Experiment}

In the light of the preceding analysis, we now take a fresh look at an
old quantum optics experiment by Ghosh and Mandel \cite{ghoshmandel}.
This experiment was among the category of first experiments showing spatial
correlation of photons. A UV laser beam was incident on a non-linear crystal
resulting in the production of a pair of photons via spontaneous parametric
down-conversion (SPDC). Such photons
are known to be entangled, and show quantum correlations. The two photons
travel in different directions at a small angle with respect to each other.
Two mirrors are used to bring the two photons together, and two detectors, in
the detection plane, detect them in coincidence (see Figure \ref{ghoshmandel}).
While a single detector
saw no interference, a coincident count of the two detectors, as a
function of their relative separation showed an interference pattern.

The visibility of interference was greater than $1/2$, which demonstrated
the non-classical nature of photons. This experiment has also been discussed
in a textbook \cite{greenstein}, and the interference pattern is believed
to be a result of non-local quantum correlation between the two photons
\cite{ghoshmandel,greenstein}.

One would notice the close similarity between
the Ghosh--Mandel experiment and our model system studying HBT effect with
entangled particles. The two detectors in the Ghosh--Mandel experiment, at
$x_1$ and $x_2$, see the photons reaching them after getting deflected
from the two mirrors. In effect they see the photons as coming from two 
spatially separated {\em virtual sources} A and B (see Figure \ref{ghoshmandel}).
With this recognition, the setup in Figure \ref{ghoshmandel} is virtually the
same as that in Figure \ref{schematic}, and 
our model system captures the essence of the Ghosh--Mandel experiment.
One might wonder if the generalized EPR state can describe the state of 
entangled photons emerging from an SPDC. In fact, the state of the SPDC
photons, produced from a Gaussian pump beam, is closely similar to the generalized
EPR state (\ref{sgepr}) if one puts $x_0$ equal to zero \cite{chan}.
The effect of the two mirrors in the Ghosh--Mandel experiment
(see Figure \ref{ghoshmandel}), is to make the two photons appear to arrive
from two spatially separated locations. Hence, the introduction of $x_0$
in (\ref{sgepr}) incorporates the effect of the two mirrors into the state
of the entangled photons.

Our analysis shows interference of entangled particles, and the visibility
is close to 1. So it reproduces the result of the Ghosh--Mandel experiment.
However, the surprising part is that the same result is
obtained when we use independent bosonic particles which are not
entangled, as shown in Section \ref{indep}. When independent bosonic
particles are used, one sees an interference in the coincidence count as
a function of the relative position of the detectors, and the visibility
is close to 1. But that is the result that is obtained in the Ghosh--Mandel
experiment too. This implies that in the Ghosh--Mandel experiment, if the
photons pairs were not entangled, the result would be the same as that for
entangled photons. Thus the effect seen in the Ghosh--Mandel experiment is
essentially the HBT effect. As the interference
in the HBT effect is independent of whether the two particles are entangled
or not, the Ghosh--Mandel experiment is not a demonstration of non-local
quantum correlation or entanglement between photons, as many seem to believe \cite[\S6.5]{greenstein}.
However, in the light of our analysis, the Ghosh--Mandel experiment was
historically the first unambiguous demonstration of interference between two
photons, which is a completely non-classical effect in itself, and probably
would not have been expected by Dirac \cite{dirac}. A word of caution might
be needed here. The HBT effect should not be naively considered a consequence
of just a physical overlap of two wave-packets of two photons \cite{shih2photon}.
It should rather be thought of as interference of two two-photon amplitudes.

\section{Conclusion}

To summarize, we have used wave-packets to study the HBT effect
in quantum particles following Bose--Einstein and Fermi--Dirac statistics,
using a recently introduced label-free analysis of indistinguishable particles.
The bunching and anti-bunching has been demonstrated through a simple
analysis. We have also analyzed the HBT effect for pairs of particles which
are entangled in position and momentum through an EPR like state. These
entangled particles also show an HBT effect which is not different from
the HBT effect in independent particles, in any noticeable way.

We have also argued that the Ghosh--Mandel experiment is essentially the
HBT effect with entangled particles. However, the interference seen in
that experiment, is not a consequence of any non-local correlation between
the two photons. Exactly the same effect would be observed if the photons
were not entangled.

\section*{Acknowledgments}
This work is inspired by the lucid exposition of the HBT effect by
N.~D. Hari Dass.
Ushba Rizwan thanks the Centre for Theoretical Physics for providing her the facility of the Centre during the course of this work.


\begin{thebibliography}{10}
\balance

\bibitem{hbt0}
Hanbury Brown R, Twiss RQ.
A new type of interferometer for use in radio astronomy. \emph{Philosophical Magazine} 1954; \textbf{45}(366): 663--682. \href{http://dx.doi.org/10.1080/14786440708520475}{\path{doi:10.1080/14786440708520475}}

\bibitem{born}
Born M, Wolf E.
\emph{Principles of Optics: Electromagnetic Theory of Propagation, Interference and Diffraction of Light}, 7th edition. Cambridge: Cambridge University Press, 2003. 

\bibitem{dirac}
Dirac PAM.
\emph{The Principles of Quantum Mechanics}, 4th edition. Oxford: Oxford University Press, 1967. \url{https://archive.org/details/DiracPrinciplesOfQuantumMechanics}

\bibitem{hbt}
Hanbury Brown R, Twiss RQ.
Correlation between photons in two coherent beams of light. \emph{Nature} 1956; \textbf{177}(4497): 27--29. \href{http://dx.doi.org/10.1038/177027a0}{\path{doi:10.1038/177027a0}}

\bibitem{fano}
Fano U.
Quantum theory of interference effects in the mixing of light from phase-independent sources. \emph{American Journal of Physics} 1961; \textbf{29}(8): 539--545. \href{http://dx.doi.org/10.1119/1.1937827}{\path{doi:10.1119/1.1937827}} 

\bibitem{hbtlight1}
Mandel L.
Quantum effects in one-photon and two-photon interference. \emph{Reviews of Modern Physics} 1999; \textbf{71}(2): S274--S282. \href{http://dx.doi.org/10.1103/RevModPhys.71.S274}{\path{doi:10.1103/RevModPhys.71.S274}} 

\bibitem{hbtlight2}
Kaltenbaek R, Blauensteiner B, Zukowski M, Aspelmeyer M, Zeilinger A.
Experimental interference of independent photons. \emph{Physical Review Letters} 2006; \textbf{96}(24): 240502. \href{http://arxiv.org/abs/quant-ph/0603048}{\path{arXiv:quant-ph/0603048}}, \href{http://dx.doi.org/10.1103/PhysRevLett.96.240502}{\path{doi:10.1103/PhysRevLett.96.240502}} 

\bibitem{hbtatom1}
Yasuda M, Shimizu F.
Observation of two-atom correlation of an ultracold neon atomic beam. \emph{Physical Review Letters} 1996; \textbf{77}(15): 3090--3093. \href{http://dx.doi.org/10.1103/PhysRevLett.77.3090}{\path{doi:10.1103/PhysRevLett.77.3090}} 

\bibitem{hbtatom2}
Schellekens M, Hoppeler R, Perrin A, Gomes JV, Boiron D, Aspect A, Westbrook CI.
Hanbury Brown Twiss effect for ultracold quantum gases. \emph{Science} 2005; \textbf{310}(5748): 648--651. \href{http://dx.doi.org/10.1126/science.1118024}{\path{doi:10.1126/science.1118024}} 

\bibitem{hbtatom3}
\"{O}ttl A, Ritter S, K\"{o}hl M, Esslinger T.
Correlations and counting statistics of an atom laser. \emph{Physical Review Letters} 2005; \textbf{95}(9): 090404. \href{http://dx.doi.org/10.1103/PhysRevLett.95.090404}{\path{doi:10.1103/PhysRevLett.95.090404}} 

\bibitem{hbtelectrons}
Neder I, Ofek N, Chung Y, Heiblum M, Mahalu D, Umansky V.
Interference between two indistinguishable electrons from independent sources. \emph{Nature} 2007; \textbf{448}: 333--337. \href{http://dx.doi.org/10.1038/nature05955}{\path{doi:10.1038/nature05955}} 

\bibitem{deb1}
Li YS, Zeng B, Liu XS, Long GL.
Entanglement in a two-identical-particle system. \emph{Physical Review A} 2001; \textbf{64}(5): 054302. \href{http://arxiv.org/abs/quant-ph/0104101}{\path{arXiv:quant-ph/0104101}}, \href{http://dx.doi.org/10.1103/PhysRevA.64.054302}{\path{doi:10.1103/PhysRevA.64.054302}} 

\bibitem{deb2}
Pa\v{s}kauskas R, You L.
Quantum correlations in two-boson wave functions. \emph{Physical Review A} 2001; \textbf{64}(4): 042310. \href{http://arxiv.org/abs/quant-ph/0106117}{\path{arXiv:quant-ph/0106117}}, \href{http://dx.doi.org/10.1103/PhysRevA.64.042310}{\path{doi:10.1103/PhysRevA.64.042310}} 

\bibitem{deb3}
Schliemann J, Cirac JI, Ku\'{s} M, Lewenstein M, Loss D.
Quantum correlations in two-fermion systems. \emph{Physical Review A} 2001; \textbf{64}(2): 022303. \href{http://arxiv.org/abs/quant-ph/0012094}{\path{arXiv:quant-ph/0012094}}, \href{http://dx.doi.org/10.1103/PhysRevA.64.022303}{\path{doi:10.1103/PhysRevA.64.022303}} 

\bibitem{deb5}
Eckert K, Schliemann J, Bru{\ss} D, Lewenstein M.
Quantum correlations in systems of indistinguishable particles. \emph{Annals of Physics} 2002; \textbf{299}(1): 88--127. \href{http://arxiv.org/abs/quant-ph/0203060}{\path{arXiv:quant-ph/0203060}}, \href{http://dx.doi.org/10.1006/aphy.2002.6268}{\path{doi:10.1006/aphy.2002.6268}} 

\bibitem{deb6}
Plastino AR, Manzano D, Dehesa JS. Separability criteria and entanglement measures for pure states of $N$ identical fermions. \emph{Europhysics Letters} 2009; \textbf{86}(2): 20005. \href{http://arxiv.org/abs/1002.0465}{\path{arXiv:1002.0465}}, \href{http://dx.doi.org/10.1209/0295-5075/86/20005}{\path{doi:10.1209/0295-5075/86/20005}} 

\bibitem{deb7}
Ghirardi G, Marinatto L.
General criterion for the entanglement of two indistinguishable particles. \emph{Physical Review A} 2004; \textbf{70}(1): 012109. \href{http://arxiv.org/abs/quant-ph/0401065}{\path{arXiv:quant-ph/0401065}}, \href{http://dx.doi.org/10.1103/PhysRevA.70.012109}{\path{doi:10.1103/PhysRevA.70.012109}} 

\bibitem{deb8}
Tichy MC, Mintert F, Buchleitner A.
Essential entanglement for atomic and molecular physics. \emph{Journal of Physics B: Atomic, Molecular and Optical Physics} 2011; \textbf{44}(19): 192001. \href{http://arxiv.org/abs/1012.3940}{\path{arXiv:1012.3940}}, \href{http://dx.doi.org/10.1088/0953-4075/44/19/192001}{\path{doi:10.1088/0953-4075/44/19/192001}}

\bibitem{deb9}
Ghirardi G, Marinatto L, Weber T.
Entanglement and properties of composite quantum systems: a conceptual and mathematical analysis. \emph{Journal of Statistical Physics} 2002; \textbf{108}(1): 49--122. \href{http://arxiv.org/abs/quant-ph/0109017}{\path{arXiv:quant-ph/0109017}}, \href{http://dx.doi.org/10.1023/a:1015439502289}{\path{doi:10.1023/a:1015439502289}} 

\bibitem{deb10}
Tichy MC, de Melo F, Ku\'{s} M, Mintert F, Buchleitner A.
Entanglement of identical particles and the detection process. \emph{Fortschritte der Physik} 2013; \textbf{61}(2--3): 225--237. \href{http://arxiv.org/abs/0902.1684}{\path{arXiv:0902.1684}}, \href{http://dx.doi.org/10.1002/prop.201200079}{\path{doi:10.1002/prop.201200079}} 

\bibitem{deb11}
Buscemi F, Bordone P, Bertoni A.
Linear entropy as an entanglement measure in two-fermion systems. \emph{Physical Review A} 2007; \textbf{75}(3): 032301. \href{http://arxiv.org/abs/quant-ph/0611223}{\path{arXiv:quant-ph/0611223}}, \href{http://dx.doi.org/10.1103/PhysRevA.75.032301}{\path{doi:10.1103/PhysRevA.75.032301}} 

\bibitem{deb12}
Reusch A, Sperling J, Vogel W.
Entanglement witnesses for indistinguishable particles. \emph{Physical Review A} 2015; \textbf{91}(4): 042324. \href{http://arxiv.org/abs/1501.02595}{\path{arXiv:1501.02595}}, \href{http://dx.doi.org/10.1103/PhysRevA.91.042324}{\path{doi:10.1103/PhysRevA.91.042324}} 

\bibitem{deb13}
Shi Y.
Quantum entanglement of identical particles. \emph{Physical Review A} 2003; \textbf{67}(2): 024301. \href{http://arxiv.org/abs/quant-ph/0205069}{\path{arXiv:quant-ph/0205069}}, \href{http://dx.doi.org/10.1103/PhysRevA.67.024301}{\path{doi:10.1103/PhysRevA.67.024301}} 

\bibitem{deb14}
Benenti G, Siccardi S, Strini G.
Entanglement in helium. \emph{European Physical Journal D} 2013; \textbf{67}(4): 83. \href{http://arxiv.org/abs/1204.6667}{\path{arXiv:1204.6667}}, \href{http://dx.doi.org/10.1140/epjd/e2013-40080-y}{\path{doi:10.1140/epjd/e2013-40080-y}} 

\bibitem{deb15}
Balachandran AP, Govindarajan TR, de Queiroz AR, Reyes-Lega AF.
Entanglement and particle identity: a unifying approach. \emph{Physical Review Letters} 2013; \textbf{110}(8): 080503. \href{http://arxiv.org/abs/1303.0688}{\path{arXiv:1303.0688}}, \href{http://dx.doi.org/10.1103/PhysRevLett.110.080503}{\path{doi:10.1103/PhysRevLett.110.080503}} 

\bibitem{deb16}
Benatti F, Floreanini R, Marzolino U.
Bipartite entanglement in systems of identical particles: the partial transposition criterion. \emph{Annals of Physics} 2012; \textbf{327}(5): 1304--1319. \href{http://arxiv.org/abs/1202.2993}{\path{arXiv:1202.2993}}, \href{http://dx.doi.org/10.1016/j.aop.2012.02.002}{\path{doi:10.1016/j.aop.2012.02.002}} 

\bibitem{deb17}
Sasaki T, Ichikawa T, Tsutsui I.
Entanglement of indistinguishable particles. \emph{Physical Review A} 2011; \textbf{83}(1): 012113. \href{http://arxiv.org/abs/1009.4147}{\path{arXiv:1009.4147}}, \href{http://dx.doi.org/10.1103/PhysRevA.83.012113}{\path{doi:10.1103/PhysRevA.83.012113}} 

\bibitem{deb18}
Benatti F, Floreanini R, Marzolino U.
Entanglement robustness and geometry in systems of identical particles. \emph{Physical Review A} 2012; \textbf{85}(4): 042329. \href{http://arxiv.org/abs/1204.3746}{\path{arXiv:1204.3746}}, \href{http://dx.doi.org/10.1103/PhysRevA.85.042329}{\path{doi:10.1103/PhysRevA.85.042329}} 

\bibitem{deb19}
Benatti F, Floreanini R, Titimbo K.
Entanglement of identical particles. \emph{Open Systems \& Information Dynamics} 2014; \textbf{21}(1--2): 1440003. \href{http://arxiv.org/abs/1403.3178}{\path{arXiv:1403.3178}}, \href{http://dx.doi.org/10.1142/s1230161214400034}{\path{doi:10.1142/s1230161214400034}} 

\bibitem{9qm}
Styer DF, Balkin MS, Becker KM, Burns MR, Dudley CE, Forth ST, Gaumer JS, Kramer MA, Oertel DC, Park LH, Rinkoski MT, Smith CT, Wotherspoon TD.
Nine formulations of quantum mechanics. \emph{American Journal of Physics} 2002; \textbf{70}(3): 288--297. \href{http://dx.doi.org/10.1119/1.1445404}{\path{doi:10.1119/1.1445404}} 

\bibitem{franco}
Lo Franco R, Compagno G.
Quantum entanglement of identical particles by standard information-theoretic notions. \emph{Scientific Reports} 2016; \textbf{6}: 20603. \href{http://arxiv.org/abs/1511.03445}{\path{arXiv:1511.03445}}, \href{http://dx.doi.org/10.1038/srep20603}{\path{doi:10.1038/srep20603}}

\bibitem{quanton}
L\'{e}vy-Leblond J-M.
Quantum words for a quantum world. In: \emph{Epistemological and Experimental Perspectives on Quantum Physics}. Greenberger D, Reiter WL, Zeilinger A (editors), Dordrecht: Springer, 1999, pp.~75--87. \href{http://dx.doi.org/10.1007/978-94-017-1454-9_5}{\path{doi:10.1007/978-94-017-1454-9_5}} 

\bibitem{epr}
Einstein A, Podolsky B, Rosen N.
Can quantum-mechanical description of physical reality be considered complete? \emph{Physical Review} 1935; \textbf{47}(10): 777--780. \href{http://dx.doi.org/10.1103/PhysRev.47.777}{\path{doi:10.1103/PhysRev.47.777}} 

\bibitem{tqajp}
Qureshi T.
Understanding Popper's experiment. \emph{American Journal of Physics} 2005; \textbf{73}(6): 541--544. \href{http://arxiv.org/abs/quant-ph/0405057}{\path{arXiv:quant-ph/0405057}}, \href{http://dx.doi.org/10.1119/1.1866098}{\path{doi:10.1119/1.1866098}}

\bibitem{tqsheeba}
Qureshi T, Shafaq S.
Wave-packet analysis of single-slit ghost diffraction. \emph{European Physical Journal Plus} 2015; \textbf{130}(8): 173. \href{http://arxiv.org/abs/1505.05559}{\path{arXiv:1505.05559}}, \href{http://dx.doi.org/10.1140/epjp/i2015-15173-6}{\path{doi:10.1140/epjp/i2015-15173-6}} 

\bibitem{ghoshmandel}
Ghosh R, Mandel L.
Observation of nonclassical effects in the interference of two photons. \emph{Physical Review Letters} 1987; \textbf{59}(17): 1903--1905. \href{http://dx.doi.org/10.1103/PhysRevLett.59.1903}{\path{doi:10.1103/PhysRevLett.59.1903}} 

\bibitem{greenstein}
Greenstein G, Zajonc AG.
\emph{The Quantum Challenge: Modern Research on the Foundations of Quantum Mechanics}, 2nd edition. Sudbury, Massachusetts: Jones \& Bartlett Publishers, 2005. 

\bibitem{chan}
Chan KW, Torres JP, Eberly JH.
Transverse entanglement migration in Hilbert space. \emph{Physical Review A} 2007; \textbf{75}(5): 050101. \href{http://arxiv.org/abs/quant-ph/0608163}{\path{arXiv:quant-ph/0608163}}, \href{http://dx.doi.org/10.1103/PhysRevA.75.050101}{\path{doi:10.1103/PhysRevA.75.050101}} 

\bibitem{shih2photon}
Pittman TB, Strekalov DV, Migdall A, Rubin MH, Sergienko AV, Shih YH.
Can two-photon interference be considered the interference of two photons? \emph{Physical Review Letters} 1996; \textbf{77}(10): 1917--1920. \href{http://dx.doi.org/10.1103/PhysRevLett.77.1917}{\path{doi:10.1103/PhysRevLett.77.1917}} 

\end{thebibliography}
\end{document}